\documentclass[12pt]{article}

\begin{document}

\begin{center}
\Large{\bf WAVE KERNEL FOR SCHR\"ODINGER OPERATOR WITH A LIOUVILLE POTENTIAL}
\end{center}
\begin{center} {\bf Yehdhih Mohamed Abdelhaye, Badahi  Mohamed 
and Mohamed Vall Moustapha}
\end{center}
\begin{abstract}{\bf In this note we give an explicit formula for the wave equation associated to the Schrodinger operator with a Liouville Potential with applications to the telegraph equation as well as
the wave equation on the hyperbolic plane.}\end{abstract}

{\bf Introduction}:Consider the following linear wave equation
$$\leqno(\alpha)
[\frac{\partial^2}{\partial X^{2}} -k^{2}e^{2X}]U(t,X)=\frac{\partial^2}{\partial t^{2}}U(t,X)$$
$$\leqno(\beta)\ \ U(0,X)=0 , \frac{\partial }{\partial t}U(0,X)=f(X) , f \in C^{\infty}_{0}(R)$$
where for $k\in R$; $\Lambda^{k}=\frac{\partial^2}{\partial X^{2}} -k^{2}e^{2X}$ is the Schrodinger operator with a Liouville potential.
It is known that $-\Lambda^{k}$ is self-adjoint positive definite
and has an absolute continuous spectrum as well as a points spectrum.
A review of the Liouville potential problems is, however, outside
the scope of this paper, and the importance of the subject for both
theory and application in mathematics and physics may be found in literature [3]. For example the purely vibrational levels of diatomic
molecules with angular momentum $l=0$
have been described by the Liouville potential for long time [7]. Another application is in string theory : the equation $(\alpha)$ is the equation of motion for a re-scaled tachyon field and is nothing but
Wheeler-de Wit equation satisfied by the macroscopic loop [6].\\
The purpose of this 
paper is to give an explicit solution of the Cauchy problem $(\alpha),(\beta)$.\\
{\bf Lemma 1}. Set:\\
$(1) Z=|k|\sqrt{2e^{X+X'}(\cosh t-\cosh (X-X'))})$\\
then we have:\\
$(2)\frac{\partial Z}{\partial X}=\frac{1}{2}Z-k^{2}e^{X+X'}\sinh (X-X') Z^{-1}$\\
$(3)\frac{\partial^{2} Z}{\partial X^{2}}=\frac{1}{4}Z-k^{2}e^{2X}Z^{-1}-k^{4}e^{2X+2X'}\sinh ^{2}(X-X') Z^{-3}$\\
$(4)\frac{\partial Z}{\partial t}=k^{2}e^{X+X'}\sinh t Z^{-1}$\\
$(5)\frac{\partial^{2} Z}{\partial t^{2}}=k^{2}e^{X+X'}\cosh t Z^{-1}-k^{4}e^{2X+2X'}\sinh ^{2}t Z^{-3}$\\
The proof of this Lemma is strait forward calculation and in consequence is left to  the reader.\\
{\bf Proposition 2}.The general solution of the wave equation $(\alpha)$
is given by:\\
$W_k(t,X,X')=aJ_0(|k|\sqrt{2e^{X+X'}(\cosh t-\cosh (X-X'))})+\\
+bY_0(|k|\sqrt{2e^{X+X'}(\cosh t-\cosh (X-X'))})$
where $a, b \in C $ and $J_0 , Y_0$ are Bessel functions of the first
and second kind respectively.\\
{Proof.} Let\\
$(6)\ D=\frac{\partial^{2}W(t,X,X')}{\partial X^{2}}-
\frac{\partial^{2}W(t,X,X'}{\partial t^{2}}=\left[(\frac{\partial Z}{\partial X})^{2}-(\frac{\partial Z}{\partial t})^{2}\right]\Phi''(Z)
+\left[\frac{\partial^{2} Z}{\partial X^{2}}-\frac{\partial^{2} Z}{\partial t^{2}}\right]\Phi'(Z)$.\\
where.\\
$(7)\ W(t,X,X')=\Phi(Z)$\\
using the formulas $(2), (3), (4)$ and $(5)$ we obtain\\
$(8)D=[\frac{1}{4}Z^{2}-k^{2}e^{X+X'}\sinh (X-X')+k^{4}e^{2X+2X'}[\sinh ^{2}(X-X')-\sinh ^{2}t]Z^{-2}]\Phi''(Z)+$\\
$[\frac{1}{4}Z-k^{2}e^{2X}Z^{-1}-k^{2}e^{X+X'}\cosh t Z^{-1}+k^{4}e^{2X+2X'}[\sinh ^{2}t-\sinh ^{2}(X-X')]Z^{-3}]\Phi'(Z)$\\
from the Bessel equation $[5]$ p.$98$\\
$Z^{2}\Phi''(Z)+Z\Phi'(Z)+(Z^{2}-\nu^{2})\Phi(Z)=0$\\
for $\nu=0$ and $Z\neq 0$ we have\\
$\Phi''(Z)=-Z^{-1}\Phi'(Z)-\Phi(Z)$.\\
replacing in $(8)$ we obtain.\\
$D=[\frac{1}{4}Z^{2}-k^{2}e^{X+X'}(\cosh (X-X')+\cosh t)Z^{-1}-
k^{4}e^{2X+2X'}(\sinh ^{2}(X-X')-\sinh^{2} t)\times$ \\
$Z^{-3}]\Phi'(Z)+
[\frac{1}{2}k^{2}e^{X+X'}(\cosh (X-X')+\cosh t)-
k^{4}e^{2X+2X'}(\sinh ^{2}(X-X')-\sinh^{2} t)Z^{-2}]\Phi(Z)$\\$+
k^{2}e^{2X}\Phi(Z)$.\\
$ (9)\ D=A(t,X,X')[2Z^{-1}\Phi'(Z)+\Phi(Z)]+k^{2}e^{2X}\Phi(Z).$\\
with\\
$A(t,X,X')={-\frac{1}{2}k^{2}e^{X+X'}(\cosh (X-X')+\cosh t)-
k^{4}e^{2X+2X'}(\sinh ^{2}(X-X')-\sinh^{2} t)Z^{-2}}$
taking into account $(1)$ and the formula
$\sinh^{2}y-\sinh^{2}z=\cosh^{2}y-\cosh^{2}z$
we obtain $A(t,X,X')= 0$, and the proof of the proposition is finished.\\
{\bf Theorem 3}. {\it The Cauchy problem $(\alpha), (\beta)$ for the wave equation with
a Liouville Potential has the unique solution given by:
$$U(t,X)=\int_{|X-X'|<t}J_0(|k|\sqrt{2e^{X+X'}(\cosh t-\cosh (X-X'))})f(X')dX'$$}
{\bf Proof }
By the proposition $2$ and the fact that the uniqueness of the solution of the problem $(\alpha),(\beta)$ is a consequence of the classical theory of hyperbolic operator; the proof of the theorem 
will be finished by showing limit conditions $(\beta)$.
\\
For this, set $\sinh(\frac{X-X'}{2})=z\sinh\frac{t}{2}$ to get\\
$(10)U(t,X)=\int^{1}_{0}J_0(2|k|\sinh \frac{t}{2}\sqrt{e^{X+X'}(1-z^{2})}
)[f(X+2\arg\sinh(z\sinh \frac{t}{2}))+f(X-2\arg\sinh(z\sinh \frac{t}{2}))]\frac{\sinh\frac{t}{2}}{\sqrt{(1+z^{2}\sinh^{2}\frac{t}{2}})}dz$.
\\
And it is not hard to see the limit conditions $(\beta)$
from $(10)$ and the formula $[5]$ p.134
$$J_\nu(x)\approx \frac{x^{\nu}}{2^{\nu}\Gamma(1+\nu)};  x\rightarrow 0.$$
We give some applications of the theorem:\\
1-Replacing $k$ by $\frac{k}{\lambda}$ ,$X$ by $\lambda X$ and $t$ by $\lambda t$ and letting $\lambda\rightarrow 0$ in $(\alpha)$,$(\beta)$ we get the solution of the Cauchy problem for the wave equation with constant potential:\\
$(a)
\left[\frac{\partial^2}{\partial X^{2}} -k^{2}\right]U(t,X)=\frac{\partial^2}{\partial t^{2}}U(t,X)$;\\
$(b)U(0,X)=0 , \frac{\partial }{\partial t}U(0,X)=f(X) , f \in C^{\infty}_{0}(R)$\\
{\bf Corollary 4}. The Cauchy problem $(a),(b)$ for the classical wave equation
with constant potential has the unique solution given by
$$U(t,X)=\int_{|X-X'|< t}J_0(|k|\sqrt{(t^{2}-(X-X')^{2})})f(X')dX'$$
Note that the telegraph equation
satisfied by the voltage or the current $v$ as a function of the 
time $t$ and the position $X$ along the cable from initial point;\\
$(a)'\frac{\partial^2 }{\partial X^{2}}v(t,X) =\frac{\partial^{2}} {\partial t^{2}}v(t,X)+(\alpha +\beta)\frac{\partial}{\partial t}v(t,X)+\alpha \beta
v(t,X)$\\
$(b)'v(0,X)=0 , \frac{\partial }{\partial t}v(0,X)=f(X) , f \in C^{\infty}_{0}(R)$\\
can be reduced to the wave equation with constant potential $(a),(b)$; where
$k=\frac{(\alpha-\beta)^{2}}{4}$;by introducing
$U=e^{((\alpha+\beta)/2)t}v$
see [1] p.192-193;695. \\
2-Consider the following Cauchy problem for the wave equation on the hyperbolic plane $[4]$\\
$(a)''Lu(t,w) =\frac{\partial ^{2}}{\partial t^{2}}u(t,X)$\\
$(b)''u(0,w)=0 , \frac{\partial }{\partial t}u(0,w)=f(w) , f \in C^{\infty}_{0}(H^{2})$\\
where $L=\Delta+\frac{1}{4}$ and $\Delta $ is the Laplace Beltrami
operator of the hyperbolic plane.\\
{\bf Corollary 5}. The Cauchy problem $(a)'',(b)''$ for the wave equation
on the hyperbolic plane has the unique solution given by
$$u(t,w)=\frac{1}{\sqrt{2}\pi}\int_{d(w,w')<t}(\cosh t-\cosh d(w,w'))^{-\frac{1}{2}}
f(w')d\mu(w')$$
where $d(w,w')$ is the geodesic distance on $H^{2}$.\\
{\bf Proof}. Let $H^{2}=\left\{w=x+iy ; x \in R , y >	0\right\}$ denote
the Poincare upper half plane with the usual hyperbolic metric:\\
$(11) ds^{2}=\frac{dx^{2}+dy^{2}}{y^{2}}$.\\
the corresponding geodesic distance $d(w,w')$ is given by\\
$(12) \cosh d(w,w')=\frac{(x-x')^{2}+y^{2}+y'^{2}}{2yy'}$\\ 
The measure associated to the metric $ds$ is\\ 
$(13) d\mu(y)=y^{-2}dx dy$.\\
and the Laplace Beltrami operator of the manifold $(H^{2},ds)$
is\\
$(14)\Delta =y^{2}(\frac{\partial^{2}}{\partial x^{2}}+\frac{\partial^{2}}{\partial y^{2}})$.\\
The Cauchy problem $(a)'',(b)''$ becomes
$$\left[y^{2}(\frac{\partial^{2}}{\partial y^{2}}+\frac{\partial^{2}}{\partial x^{2}})+\frac{1}{4}\right]u(t,w)
=\frac{\partial^{2} u(t,w)}{\partial t^{2}}$$
$$u(0,w)=0 , \frac{\partial u(0,w)}{\partial t}=f(w) , f \in C_{0}^{\infty}(H^{2})$$
The Fourier transform with respect to the variable $x$ gives\\
$\left[y^{2}\frac{\partial^{2}}{\partial y^{2}}-k^{2}y^{2}+\frac{1}{4}\right]\hat{u}(t,k,y)
=\frac{\partial \hat{u}(t,k,y)}{\partial t^{2}}$\\
$\hat{u}(0,k,y)=0 , \frac{\partial \hat{u}(0,k,y)}{\partial t}=\hat{f}(k,y) , f \in C^{\infty}_{0}(H^{2})$.\\                        
Set\\
$(15) \hat{u}(t,k,y)=y^{\frac{1}{2}}v(t,k,y)$\\
$(16)y=e^{X}$\\
$(17) v(t,k,y)=U(t,k,X)$\\
We obtain\\
$[\frac{\partial^2}{\partial X^{2}} -k^{2}e^{2X}]U(t,k,X)=\frac{\partial^{2}}{\partial t^{2}}U(t,k,X)$\\
$U(0,k,X)=0 , \frac{\partial }{\partial t}U(0,k,X)=g(k,X) , f \in C^{\infty}_{0}(H^{2})$;\\
with\\
$(18)\ g(k,X)=e^{-\frac{X}{2}}\hat{f}(k,e^{X})$\\
Using the theorem $3$ we get
$$U(t, k, X)=\int_{|X-X'|< t}J_0(|k|\sqrt{2e^{X+X'}(\cosh t-\cosh (X-X'))})g(k, X')dX'$$
From $(18)$,$(17)$,$(16)$ and $(15)$
we obtain\\
$u(t, w)=\frac{\sqrt{yy'}}{2\pi}\int^{\infty}_{-\infty}\int_{|\ln \frac{y}{y'}|< t}\int^{\infty}_{-\infty}e^{-ik(x-x')}\times$\\
$$J_0(|k|\sqrt{2yy'\cosh t-y^{2}-y'^{2})})dk f(x',y')\frac{dx'dy'}{y'^{2}}$$
$u(t,w)=\frac{\sqrt{yy'}}{\pi}\int^{\infty}_{-\infty}\int_{|\ln \frac{y}{y'}|< t}\int^{\infty}_{0}\cos k(x-x')\times$
$$J_0(k\sqrt{2yy'\cosh t-y^{2}-y'^{2})})dk f(x',y')\frac{dx'dy'}{y'^{2}}$$
and hence by the formula $7.165$ p.$182$ of $[2]$,
$$\int^{\infty}_{0}\cos ut J_{0}(at)dt= \left\{\begin{array}{cc}(a^{2}-u^{2})^{-\frac{1}{2}}& \mbox{if $u<a$}\\
0&\mbox{$u>a$}
\end{array}\right.$$
and the formulas $(12)$ and $(13)$ we get the proof of the corollary.\\
\begin{center}\bf 
References
\end{center}
${\bf[1]}$-R. Courant and D. Hilbert; Methods of Mathematical Physics, Volume II,Interscience Publisher
1962.\\
${\bf [2]}$-V. Ditkine et A. Proudnikov; Transformations
integrales et calcul op\'erationel; Traduction francaise
Edition MIR 1978.\\
${\bf [3]}$-N. Ikeda and H. Matsumoto; Brownien motion
one the hyperbolic plane and Selberg trace formula; J. Funct. Anal. 163(1999), 63-110.\\
${\bf [4]}$-A. Intissar et M. V. Ould Moustapha; Solution explicite de l'\'equation des ondes dans un espace sym\'etrique de type non compact de rang 1;C.R.Acad. Sci. Paris 321(1995)77-81.\\
${\bf [5]}$-N. N. Lebedev; Special Functions
and their applications; Dover Publications INC New York 1972.\\
${\bf [6]}$-Miao Li; Some remarks on tachyon action
in $2d$ string theory; arXiv:hep:th/9212061 vl 9Dec 92.\\
${\bf [7]}$-H. Tasseli; Exact solutions for vibrational
Levels of Morse potential; J.Phys. A: Math.Gen.
31(1998)779-788.\\
\begin{flushleft}Department of Mathematics\\
College of Arts and Sciences at Al Qurayat   \\
 Al-Jouf University Kingdom of Saudi Arabia  \\
and\\
D\'epartement de Math\'ematiques et Informatique\\
 Facult\'e des Sciences et Techniques\\
Universit\'e de  Nouakchott Al-Aasriya\\
Nouakchott-Mauritanie\\
Email: mohamedvall.ouldmoustapha230@gmail.com
\end{flushleft} 
\end{document}